\documentclass[journal,twoside,web]{ieeecolor}
\usepackage{tmi}
\usepackage{cite}
\usepackage{amsmath,amssymb,amsfonts}
\usepackage{algorithmic}
\usepackage{graphicx}
\usepackage{textcomp}
\usepackage{epstopdf}
\usepackage{multirow}
\usepackage{caption}
\usepackage{subfigure}
\usepackage{booktabs}
\let\labelindent\relax
\usepackage{enumitem}
\usepackage[table]{xcolor}
\usepackage{tabularx,ragged2e}
\usepackage{url}

\usepackage{algorithm}
\usepackage{algorithmic}

\def\BibTeX{{\rm B\kern-.05em{\sc i\kern-.025em b}\kern-.08em
		T\kern-.1667em\lower.7ex\hbox{E}\kern-.125emX}}
\begin{document}
	\title{Harmonizing Pathological and Normal Pixels for Pseudo-healthy Synthesis}
\author{Yunlong~Zhang, Xin~Lin, Yihong~Zhuang, Liyan~Sun, Yue~Huang, \\Xinghao~Ding, Guisheng~Wang, Lin~Yang, and Yizhou~Yu, \IEEEmembership{Fellow,~IEEE}%
		\thanks{The work is supported in part by National Key Research and Development Program of China (No. 2019YFC0118101), in part by the National Natural Science Foundation of China under Grants 82172033, 52105126, U19B2031, 61971369, in part by China Postdoctoral Science Foundation (No. 2021M702726), in part by Science and Technology Key Project of Fujian Province(No. 2019HZ020009).}
		\thanks{Yunlong Zhang was with the School of Informatics, Xiamen University, Xiamen 361005, China. He is now with the School of Engineering, Westlake University, Hangzhou, 310012, China}
        \thanks{Xin Lin, Yihong Zhuang, Yue Huang, and Xinghao Ding are with the School of Informatics, Xiamen University, Xiamen 361005, China, and the Institute of Artificial Intelligence, Xiamen University, Xiamen, 361005, China (corresponding author: Yue Huang)}
        \thanks{Liyan Sun is with the School of Electronic Science and Engineering, Xiamen University, Xiamen 361005, China}
		\thanks{Guisheng Wang is with the Department of Radiology, the Third Medical Centre, Chinese PLA General Hospital, Beijing, China}
		\thanks{Yizhou Yu is with the Deepwise AI Laboratory, Beijing 100125, China and the Department of Computer Science, The University of Hong Kong, Hong Kong}
		\thanks{Lin Yang is with the School of Engineering, Westlake University, Hangzhou 310012, China}}
	
	\maketitle
	
	\begin{abstract}
		Synthesizing a subject-specific pathology-free image from a pathological image is valuable for algorithm development and clinical practice. In recent years, several approaches based on the Generative Adversarial Network (GAN) have achieved promising results in pseudo-healthy synthesis. However, the discriminator (i.e., a classifier) in the GAN cannot accurately identify lesions and further hampers from generating admirable pseudo-healthy images. To address this problem, we present a new type of discriminator, the segmentor, to accurately locate the lesions and improve the visual quality of pseudo-healthy images. Then, we apply the generated images into medical image enhancement and utilize the enhanced results to cope with the low contrast problem existing in medical image segmentation. Furthermore, a reliable metric is proposed by utilizing two attributes of label noise to measure the health of synthetic images. Comprehensive experiments on the T2 modality of BraTS demonstrate that the proposed method substantially outperforms the state-of-the-art methods. The method achieves better performance than the existing methods with only 30\% of the training data. The effectiveness of the proposed method is also demonstrated on the LiTS and the T1 modality of BraTS. The code and the pre-trained model of this study are publicly available at \url{https://github.com/Au3C2/Generator-Versus-Segmentor}.
	\end{abstract}
	
	
	\section{Introduction}
	\label{sec:introduction}
	Pseudo-healthy synthesis is defined as synthesizing a subject-specific pathology-free image from a pathological image \cite{bowles2016pseudo,xia2020pseudo}. Generating such images has been demonstrated to be valuable for a variety of tasks in medical image analysis \cite{xia2020pseudo}, such as segmentation \cite{bowles2017brain,ye2013modality,sun2020adversarial,andermatt2018pathology,bowles2016pseudo,du2021disease}, detection \cite{tsunoda2014pseudo}, and assisting doctors with diagnosis by comparing the pathological and pseudo-healthy images \cite{baumgartner2018visual,sun2020adversarial,tsunoda2014pseudo}. \textit{By definition, a perfect pseudo-healthy image should maintain healthiness (i.e., synthesizing healthy-like appearances) and subject identity (i.e., belonging to the same subject as the input). Both attributes are essential and indispensable. The former attribute is self-explanatory,  while the latter one is also considerable since generating another healthy image is meaningless.}

\begin{figure}
		\centering
		\subfigure[Input]{\includegraphics[width=0.48\columnwidth]{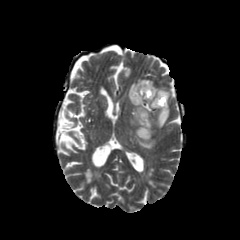}\label{fig11}}
		\subfigure[Label]{\includegraphics[width=0.48\columnwidth]{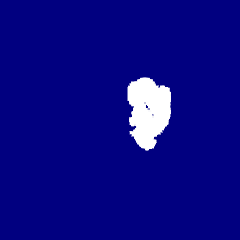}\label{fig12}}\\
		\subfigure[Classifier]{\includegraphics[width=0.48\columnwidth]{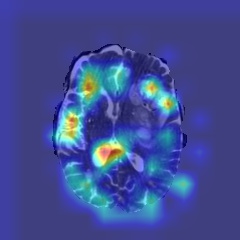}\label{fig13}}
		\subfigure[Segmentor]{\includegraphics[width=0.48\columnwidth]{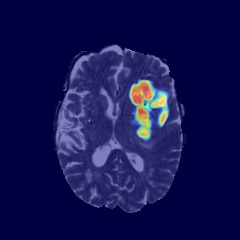}\label{fig14}}
		\caption{'Visual explanations'  of the classifier and the segmentor. (a-d) represent the input, tumor annotation, class activation maps of the classifier generated by Grad-CAM \cite{selvaraju2017grad}, and class activation maps of the segmentor generated by and Seg-Grad-CAM \cite{vinogradova2020towards}, respectively.
		}\label{fig1}
	\end{figure}

	Pseudo-healthy image synthesis is an ill-posed inverse procedure since there exists a multitude of healthy-looking solutions for a pathological input. To tackle this inverse problem, several GAN-based methods \cite{baumgartner2018visual,sun2020adversarial,xia2020pseudo} \textcolor{black}{have been}  presented. The basic architecture in these methods includes a generator and a discriminator. The generator is an encoder-decoder architecture trained to translate pathological images into corresponding healthy-looking ones, whereas the discriminator competes against the generator and aims to differentiate the synthetic and healthy images by a two-way classifier. However, choosing the classifier as the discriminator has the shortcoming that lesions cannot be accurately located (see Fig. \ref{fig13}) due to \textcolor{black}{the following reasons}: (1) The visual explanations of the classifier involve healthy regions, which will further lead to falsely erasing the subject identity. (2) The highlighted explanations of the classifier cannot cover the entire tumor region, which easily causes the pathology remains in the synthetic images.
	
	To address the problem of inaccurate localization, we choose to use a segmentor as the discriminator. The segmentor identifies pathological regions more accurately than the classifier (e.g., the 'tumor' explanation accurately highlights \textcolor{black}{the} tumor regions in Fig. \ref{fig14}). Therefore, both the subject identity and healthiness can be simultaneously guaranteed by keeping the healthy pixels and transforming the pathological ones, respectively. The former is implemented by a visual consistency loss, and the latter is achieved by the adversarial training between the generator and the segmentor.
	
	During the process of alternatively training the generator and the segmentor, the pathological pixels are gradually transformed into healthy-looking pixels. As a result, these healthy-looking pixels (i.e, should be labeled as 'no tumor') are falsely labeled as 'tumor' when training the segmentor, which results in the poor generalization of the segmentor and further hampers the entire training process. To resist false labels as much as possible, we propose a difference-aware loss to improve the generalization ability of the segmentor by muting those deceptively healthy-looking pixels.

	Moreover, we also develop one downstream task of pseudo-healthy synthesis that increases the contrast between lesions and normal tissue. It is well-known that the low contrast, especially for lesions, is a nasty nature of medical images that may result in miss detection of low-contrast anatomies in medical image segmentation. Since the proposed method can effectively disentangle a disease image into its two components, \textcolor{black}{the healthy and the pathological parts that only contain healthy tissue information and pathological information, respectively.} We design an image enhancement technique that highlights lesions by simply adding the extracted pathological signals into the inputs. Then, this technique is applied to alleviate the low contrast problem for medical image segmentation.
	
	Lacking a good quantitative metric for measuring the healthiness is one of the major barriers in pseudo-healthy synthesis. Recently, Xia et al. \cite{xia2020pseudo} proposed a metric using a pre-trained segmentor. However, this metric \textcolor{black}{has} \textit{poor stability (i.e., fluctuating drastically at multiple trials) and vulnerability (i.e., predicting falsely when synthetic pathology slightly deviates from the true pathology but is apparently abnormal)}. Besides, the subjective assessment \textcolor{black}{was} viewed as the gold standard to ultimately determine the healthiness \cite{xia2020pseudo}. However, it is time-consuming, costly, and subject to inter- and intra-observer variability, and hence also deviates from the reproducibility. Inspired by two attributes of the label noise that fitting incorrect labels requires more time \cite{zhang2016understanding} and the higher learning rate hampers the memorization of false labels \cite{Tanaka2018JointOF}, we propose a reliable metric for measuring the healthiness by estimating the convergence speed.
	
	\textcolor{black}{In order to evaluate the effectiveness of the proposed approach,} we conduct extensive experiments on the T2 modality of BraTS dataset \cite{bakas2017advancing,menze2014multimodal,bakas2018identifying}.  Furthermore, we also present a part of results on the T1 modality of BraTS and the LiTS datasets \cite{bilic2019liver} to demonstrate that the adaptive capability of our method to other modalities and organs.
	
	This work is a significant extension of our prior conference paper \cite{zhanggvs}. The main contributions are summarized as follows, (1) {\color{black}This paper further clearly and intuitively unravels the motivation of the \textcolor{black}{Generative versus Segmentor (GVS)} by using the class activation mapping technique.} (2) This paper proposes a healthiness metric by addressing the instability of the $\mathbb{S}_{dice}$ in the conference version. Then, the reliability of the improved metric is validated by \textcolor{black}{multiple experiments}. Besides, this paper introduces the subjective assessments to further evaluate the effectiveness of the proposed method. (3) This paper develops the application of pseudo-healthy synthesis for image enhancement. The enhanced images not only contribute to identifying lesions in visual but also improve lesion segmentation performance.
	
	\section{Related Works}
	\label{sec:RelatedWorks}
	\subsection{Pseudo-healthy synthesis}
	Recently, pseudo-healthy synthesis \textcolor{black}{has attracted} considerable attention in the medical image analysis community because of its potential for downstream tasks \cite{bowles2017brain,ye2013modality,sun2020adversarial,andermatt2018pathology,tsunoda2014pseudo,chen2018unsupervised,baumgartner2018visual,xia2020pseudo}. The related work is divided into two main categories, pathology-deficiency (i.e., only providing healthy images in the training phase) \cite{chen2018unsupervised,sato2018primitive,schlegl2019f} and pathology-sufficiency based methods (i.e., possessing plenty of pathological and healthy images in the training phase) \cite{baumgartner2018visual,sun2020adversarial,xia2020pseudo}.
	
	The pathology-deficiency based methods \cite{baur2018deep,zimmerer2019unsupervised,zimmerer2018context,chen2018unsupervised,pawlowski2018unsupervised,Baur2020ScaleSpaceAF,Baur2020SteGANomalyIC,Nguyen2020UnsupervisedRA} were always closely associated with unsupervised anomaly detection/segmentation \cite{baur2021autoencoders}, which aimed to learn normative distributions by learning to compress and recover healthy anatomies in the training phase. In the subsequent testing stage, pathological images were first compressed to the latent space. Then, pseudo-healthy images were reconstructed from latent representations based on the assumption that the obtained latent representations were close to the latent representations of pseudo-healthy images. However, the assumptions of these methods were too idealistic \cite{chen2018unsupervised,schlegl2019f}. Actually, the healthiness and the subject identity were not guaranteed due to the difficulty in finding the optimal latent representations, corresponding to pseudo-healthy images when compressing the pathological images into latent space.
	
	The pathology-sufficiency based methods \cite{baumgartner2018visual,sun2020adversarial,xia2020pseudo} tackled pseudo-healthy synthesis from the viewpoint of image translation. In the training phase, these methods introduced pathological images alongside corresponding image-level \cite{baumgartner2018visual} or pixel-level \cite{sun2020adversarial,xia2020pseudo} pathological annotations to learn an image translation process of mapping pathological images to pseudo-healthy images.  Baumgartner et al. \cite{baumgartner2018visual} proposed a basic GAN-based scheme, that was composed of a generator and a discriminator. The generator \textcolor{black}{was} trained to synthesize healthy-looking appearances and keep the subject identity at the same time, whereas the discriminator aimed to differentiate the synthetic images from the unpaired healthy images. This method only used the image-level annotations and \textcolor{black}{was not able to} accurately translate pathological pixels and keep the healthy ones. To alleviate these issues, PHS-GAN \cite{xia2020pseudo} and ANT-GAN \cite{sun2020adversarial} introduced pixel-level annotations. \textcolor{black}{Both methods} were variants of Cycle-GAN \cite{zhu2017unpaired}. \textcolor{black}{Specifically,} the PHS-GAN considered the one-to-many problem and disentangled the information of pathology from what seems to be healthy, and pixel-level labels were used to extract the location and shape of pathology. In the process of applying Cycle-GAN to pseudo-healthy synthesis, the ANT-GAN proposed the shortcut to simplify the optimization and the masked L2 loss to better preserve the normal regions.
	
	Our experimental setting is identical to the work by Sun et al. \cite{sun2020adversarial} and Xia et al. \cite{xia2020pseudo}. However, our motivation is significantly different from theirs. They tried to translate the pathological images into healthy-looking appearances. In comparison, the proposed method further explicitly \textcolor{black}{utilizes} the information inside the appearance differences between healthy and pathological regions and \textcolor{black}{tries} to generate pathology-free images by making up such differences until achieving harmony between them.
	
	\begin{figure*}[htbp]
		\centering
		\includegraphics[width=\textwidth]{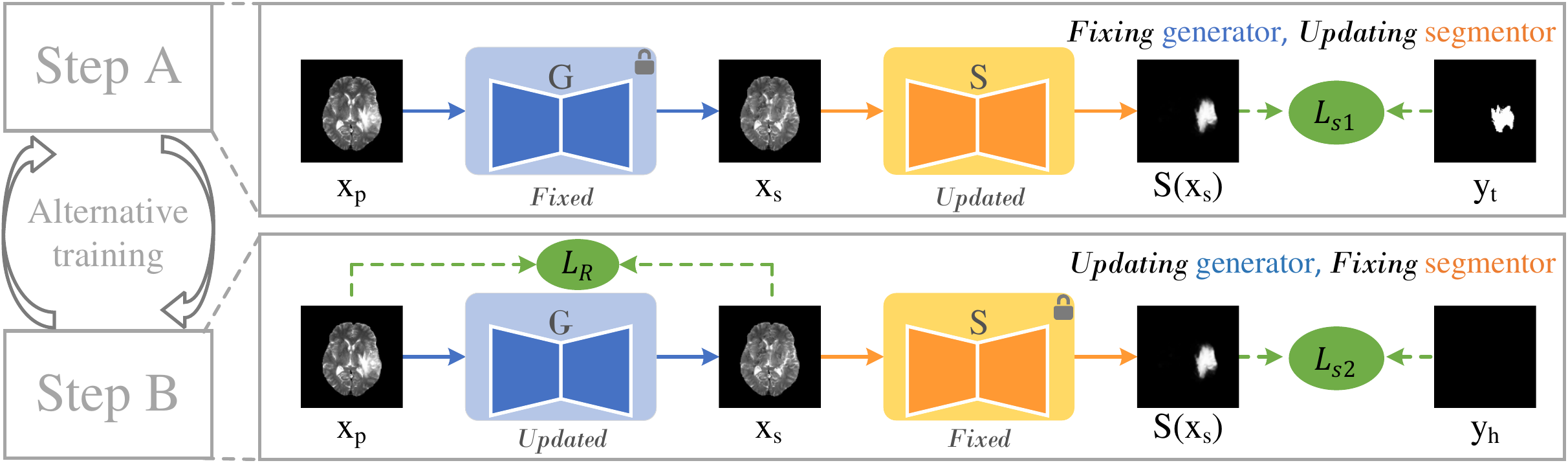}
		\caption{Training workflow. The model is optimized by iteratively alternating Step A and Step B (left chart). In Step A (right top), we fix the generator $\mathbf{G}$ and update the segmentor $\mathbf{S}$ with $L_{s1}$. In Step B (right bottom), we fix the segmentor $\mathbf{S}$ and update  the generator $\mathbf{G}$ with $L_{s2}+\lambda L_R$.}
		\label{fig2}
	\end{figure*}

	\subsection{Adversarial training}
		The idea of adopting the segmentor as the discriminator is inspired by the extensive applications of adversarial training \cite{goodfellow2014generative}. The discriminator of the original GAN \cite{goodfellow2014generative} was used to differentiate true or counterfeit images. In domain adaptation, DANN \cite{ganin2016domain}  adopted a discriminator to distinguish the data sampled from the source or the target domain. In the research on the adversarial attack,  the discriminator is a classifier to sort an example into a corresponding class \cite{goodfellow2014explaining,miyato2018virtual}. In image translation tasks, the discriminator is also a classifier to detect the high-level structured differences between the translated and real images \cite{isola2017image,zhu2017unpaired}.  Recently, Naveen et al. \cite{minderer2020automatic} applied adversarial training into self-supervised learning and adopted a classifier as the discriminator to predict pretext labels. Compared with the related work using diverse classifiers as discriminators, this paper further develops the paradigm of adversarial training and extends the discriminator to pixel-level dense prediction tasks.
		
		\subsection{Medical Image Enhancement}
		
		It is well-known that the low contrast is an intrinsic nature of medical images, which will hinder the clinical decision-making and the downstream analysis tasks (e.g., segmentation, detection). To alleviate this issue, image enhancement has been widely studied in the field of medical imaging \cite{litjens2017survey}. Singh et al. \cite{singh2019histogram} and Wang et al. \cite{wang2008fast} improved the histogram equalization, the most basic enhancement technique, to improve the visual quality of  low radiance retinal images and multiple types of medical images. Frosio et al. \cite{frosio2005enhancing}  enhanced the digital cephalic radiography with mixture models and local gamma correction. Hamghalam et al. \cite{hamghalam2020high} proposed an image-to-image translation technique to generate synthetic high tissue contrast (HTC) images and used the enhanced images to improve the segmentation performance. However, most of the existing studies did not consider the context. Hence, they can easily fail when encountering complex contexts. For example, the cerebrospinal fluid and the tumor both exhibit high signal intensity on T2-weighted images of  BraTS dataset. The existing methods cannot identify them and increase the contrast between them, which is harmful to identify the lesion after enhancement. Compared with the existing methods, the proposed enhancement method considers the context of images and can adaptively enhance the contrast between lesions and normal tissues.

	\section{Methods}
	\label{sec:Methods}
	{\color{black}The proposed Generative versus Segmentor (GVS) for pathology-sufficiency pseudo-healthy synthesis with pixel-level labeling is introduced in Sec.  \ref{subsection:flowchart} and \ref{subsection:segmentor}. Then, in Sec. \ref{subsection:enhancement}, we apply synthetic images to medical image enhancement. Furthermore, the enhanced images are used to help low-contrast medical image segmentation.}
	
	Assuming that a set of pathological images $\{x_p\}$ with their pixel-level lesion annotations $\{y_t\}$ are given. Our goal is to train a generator $\mathbf{G}$ that can translate the pathological image $x_p$ into a corresponding synthetic image $x_s$ with superior healthiness and subject identity.

	\subsection{Basic GVS flowchart}
	\label{subsection:flowchart}
	The training workflow of the proposed GVS is shown in Fig. \ref{fig2}. The generator gradually synthesizes the healthy-looking images by iteratively alternating Steps A and B. The specific steps are described as follows.
	
	\noindent\textbf{Step A.} As shown in Step A of Fig. \ref{fig2}, we fix the generator $\mathbf{G}$ and update the segmentor $\mathbf{S}$ to detect the lesions in the synthetic images. {\color{black}To this end, the segmentation loss used to train the segmentor is defined as:}
	\begin{equation}
		\mathcal{L}_{s1} = \mathcal{L}_{ce}(\mathbf{S}(x_s)), y_t),
	\end{equation}
	where $L_{ce}$ denotes the cross-entropy loss. Otherwise, the $y_t$ are binary labels, \textcolor{black}{where 0 and 1 represent the normal and the pathological regions, respectively.}
	
	\noindent\textbf{Step B.} In this step, we fix the segmentor $\mathbf{S}$ and update the generator $\mathbf{G}$, aiming to remove the lesions and preserve the subject identity of the pathological images.
	
	Firstly, it is expected that the generator $\mathbf{G}$ can synthesize healthy-looking images. To achieve this, the generator is trained to deceive the segmentor by compensating the appearance difference between pathological and healthy regions. \textcolor{black}{Specifically}, another segmentation loss is adopted as follows:
	\begin{equation}\label{eq2}
		\mathcal{L}_{s2} = \mathcal{L}_{ce}(\mathbf{S}(\mathbf{G}(x_p))), y_h),
	\end{equation}
	where $y_h$ denotes a zero matrix with the same size as $y_t$.
	
	Secondly, the synthetic images should keep the same subject identity with inputs. Hence, the residual loss proposed in the existing pseudo-healthy synthesis method \cite{sun2020adversarial} is used:
	\begin{equation}\label{eq3}
		\mathcal{L}_R = \mathcal{L}_{mse}(x_p,  \mathbf{G}(x_p)),
	\end{equation}
	where $\mathcal{L}_{mse}$ denotes the pixel-wise $\mathcal{L}_2$ loss.  The total training loss of  $\mathbf{G}$ is:
	\begin{equation}
		\mathcal{L}_{G} = \mathcal{L}_{s2} + \lambda \mathcal{L}_R,
	\end{equation}
	where $\lambda$ denotes a hyperparameter that trades off the healthiness against subject identity, and $\lambda > 0$.
	
	During the iterative training, the segmentor $\mathbf{S}$ and the generator $\mathbf{G}$ compete with each other. The segmentor $\mathbf{S}$ tries to detect the differences between normal and pathological regions, whereas the generator $\mathbf{G}$ tries to compose them. Eventually, the generator $\mathbf{G}$ bridges the gap between the pathological and normal regions and synthesizes healthy-looking images.

	\begin{figure*}
		\centering
		\includegraphics[width=\textwidth]{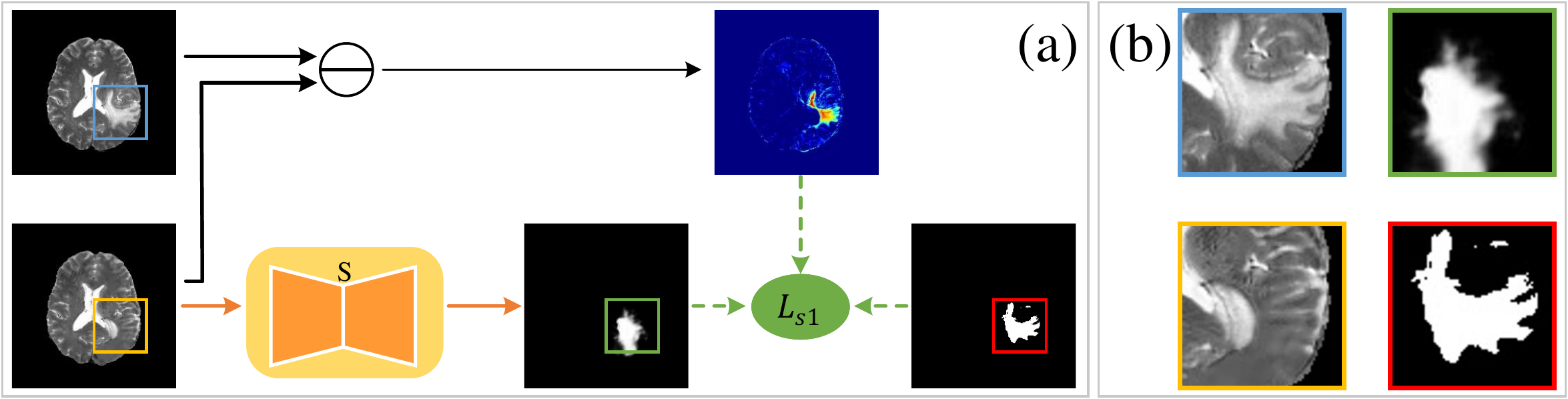}
		\caption{(a) Framework of the pixel-level weighted cross-entropy loss. (b) The blue, yellow, green, and red boxes denote the pathological image, synthetic image, prediction, and lesion annotation, respectively.
		}\label{fig4}
	\end{figure*}
	
	\subsection{Difference-aware loss}
	\label{subsection:segmentor}
	In this section, the generalization ability of the segmentor is further considered. During {\color{black}the training process}, the pathological regions are gradually transformed into normal regions. As shown in the yellow box of Fig. \ref{fig4}(b), a major part of the pathological region has been well transformed, so these pixels should be labeled as 'no tumor'. However, the basic GVS still considers these pixels as 'tumor' when training the segmentor, which misguides the segmentor and strongly harms its generalization ability \cite{zhang2016understanding}.
	The green box in Fig. \ref{fig4}(b) shows that the predictions of the segmentor severely deviate from the labels, suggesting the poor generalization ability of the segmentor.  To overcome this challenge, we adopt an easy yet effective strategy that is muting the well-transformed pixels. It is discovered that the difference maps between inputs and synthetic outputs can reflect how well the pixels are transformed. That is, the substantial differences denote the good transformation, \textcolor{black}{while the minor differences denote the poor transformation} (rf. Fig. 3(a)). Hence, \textcolor{black}{the difference maps are utilized as indicators} to measure the transformation degree and a difference-aware loss is proposed for alleviating the overfitting when training the segmentor, which is defined as:
	\begin{equation}
		\mathcal{L}_{wce} = \frac{1}{N}\sum_{i=1}^{N}w(i)y_{t}(i)log(\mathbf{S}(\mathbf{G}(x_p))(i)),
	\end{equation}
	where $N$ denotes the number of pixels.  The weights $w$ associated with difference maps are defined as:
	
	\begin{equation}
		w =  \left\{
		\begin{array}{rcl}
			0.1,       &      & 1 - m< 0.1,\\
			1 - m,       &      & \text{Otherwise},
		\end{array} \right.
	\end{equation}
	where $m = \text{Normalization}(x_p - \mathbf{G}(x_p))$ denotes the normalized difference map.
	In this work, $w[w < 0.1] = 0.1$ because the minimum value does not represent perfect transformation, and it is necessary to keep a subtle penalty. The complete GVS is proposed by upgrading the segmentation loss $\mathcal{L}_{s1}$ in Equation \ref{eq2} to the difference-aware loss $\mathcal{L}_{wce}$ in Equation 5.

	\subsection{Lesion contrast enhancement and downstream segmentation}
	\label{subsection:enhancement}
	{\color{black}After the training process, the proposed GVS can effectively disentangle a disease input into healthy (i.e., only containing healthy tissue information) and pathological (i.e., only containing pathological information) parts. Our enhancement is implemented by adding the extracted pathological part into the input, which is formulated as:
	\begin{equation}
		x_{en} = x_p + \alpha * (x_s - x_p),
	\end{equation}
	where $\alpha$ denotes the degree of enhancement and $x_s - x_p$ represents the pathological residue between the disease and pseudo-healthy images. The enhanced images $x_{en}$ increase the intensity of lesions while keeping the normal tissues well when the $x_s - x_p$ only contains the pathological information. The proposed GVS can achieve this well and thus can effectively enhance the contrast between the lesions and normal tissues.
	
	Recently, Hamghalam et al. \cite{hamghalam2020high} revealed that increasing the contrast between tissues can effectively improve the generalization ability of the segmentation task on the BraTS dataset. Inspired by this, we apply the enhanced images to improve lesion segmentation performance. Specifically, the downstream segmentor $S_D$\footnote{To distinguish the segmentor in the downstream segmentation and the GVS, the downstream segmentor $S_D$ is adopted in the downstream segmentation. similarly, in Sec. \ref{sec4}, the evaluation segmentor $S_E$ denotes the segmentor using in the evaluation process.} is trained on the enhanced images. In the subsequent test phase, test samples are also enhanced before being fed into the segmentor.}
	
	\begin{figure}
		\centering
		\subfigure[Inputs]{\begin{minipage}[b]{0.24\linewidth} \includegraphics[width=\linewidth]{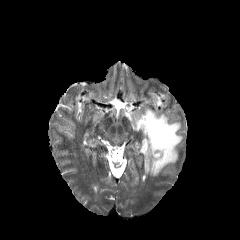}\vspace{1pt} \includegraphics[width=\linewidth]{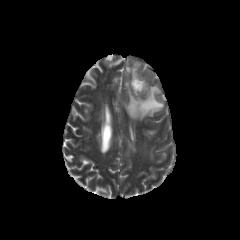}\vspace{1pt}\end{minipage}\label{fig61}}\subfigure[Counterfeit images]{\begin{minipage}[b]{0.24\linewidth} \includegraphics[width=\linewidth]{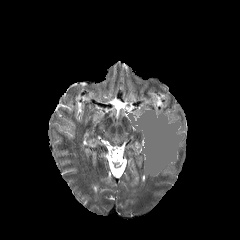}\vspace{1pt} \includegraphics[width=\linewidth]{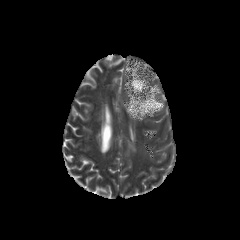}\vspace{1pt}\end{minipage}\label{fig62}}\subfigure[Predictions]{\begin{minipage}[b]{0.24\linewidth} \includegraphics[width=\linewidth]{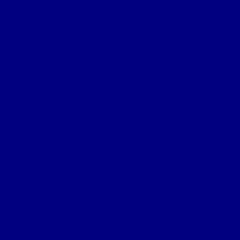}\vspace{1pt} \includegraphics[width=\linewidth]{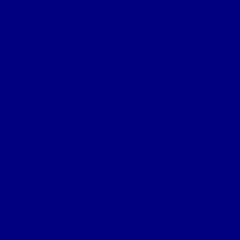}\vspace{1pt}\end{minipage}\label{fig63}}\subfigure[Labels]{\begin{minipage}[b]{0.24\linewidth} \includegraphics[width=\linewidth]{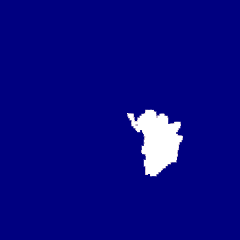}\vspace{1pt} \includegraphics[width=\linewidth]{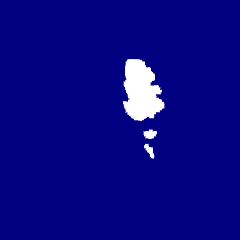}\vspace{1pt}\end{minipage}\label{fig64}}
		\caption{Two types of counterfeit images fooling the pre-trained segmentor in literature \cite{xia2020pseudo}. Four columns from left to right represent inputs, counterfeit images, the predictions of counterfeit images, and lesion annotations, respectively. The counterfeit image in the first row is generated by filling the pathological regions with the average value of normal tissues. The second row adds the Gaussian noise with zero mean and 0.2 covariance in the pathological regions. {\color{black}The pre-trained segmentor cannot detect the abnormalities existing in both counterfeit images.}}\label{fig6}
	\end{figure}

	\begin{figure*}
		\centering
		\subfigure["Healhiness"\cite{xia2020pseudo}]{\includegraphics[width=0.25\textwidth]{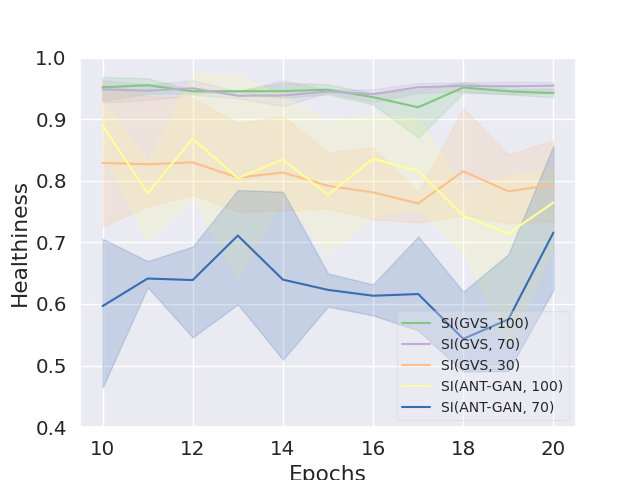}\label{fig31}}\subfigure[Dice curve (lr=0.0001)]{\includegraphics[width=0.25\textwidth]{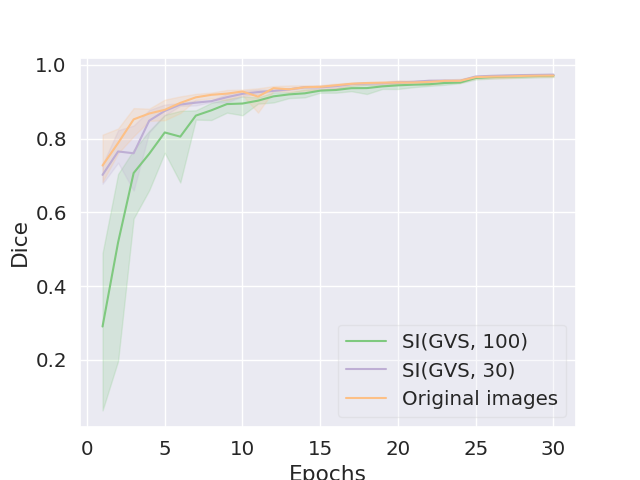}\label{fig32}}\subfigure[Dice curve (lr=0.1)]{\includegraphics[width=0.25\textwidth]{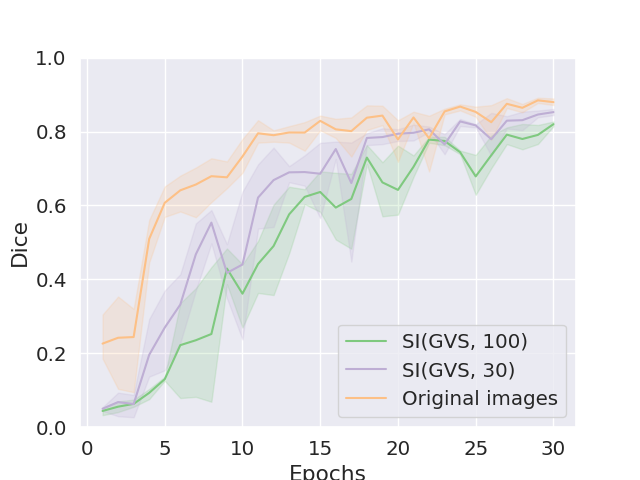}\label{fig33}}\subfigure[A-Dice]{\includegraphics[width=0.25\textwidth]{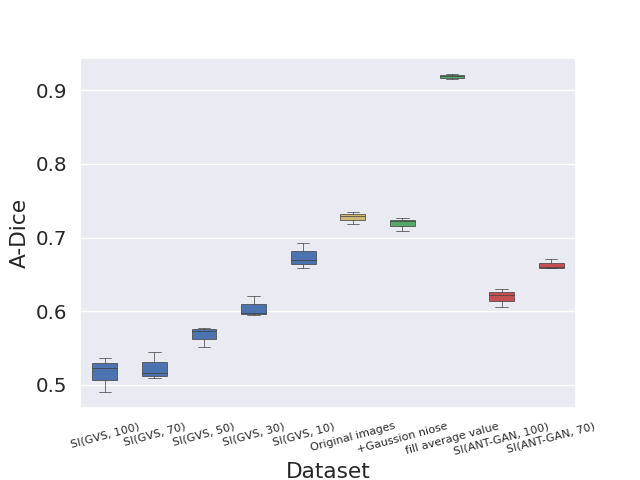}\label{fig34}}
		\caption{{\color{black}(a) The "Healthiness" proposed in literature \cite{xia2020pseudo} fluctuates drastically at various epochs and runtimes, illustrating its instability. (b-c) Dice scores on the training data when $lr=0.0001$ and $lr=0.1$, respectively. Increasing the learning rate enlarges the gap of convergence speed. (d) The A-Dice scores evaluated on different synthetic images. The "SI(Method, $a$)" in the figure represents synthetic images generated by "Method" (e.g., GVS) which is trained on $a\%$ of training data. The proposed A-Dice accurately reflects the relative order for different kinds of images and fluctuates slightly.}}\label{fig3}
	\end{figure*}

	\section{Measuring Healthiness} \label{sec4}
	{\color{black}In this section, we analyze the "Healthiness" metric proposed by Xia et al. \cite{xia2020pseudo} in the subsequent second paragraph. In the third paragraph, we introduce the motivation of the proposed healthiness metric. In the fourth paragraph, we propose a new metric A-Dice to measure the healthiness based on $\mathbb{S}_{dice}$. In the last two paragraphs, the reliability of the proposed metric is validated by analysis and experiments.}
	
    \textcolor{black}{In the evaluation process of "Healthiness", a segmentor is first pre-trained} to estimate pathology in images. Then the pre-trained segmentor is used to assess pathology within synthetic images by checking how large the estimated pathological areas are. \textcolor{black}{Further implementation details can be found in Sec. 4.4 of literature} \cite{xia2020pseudo}. This process \textcolor{black}{seems} reasonable but cannot guarantee accurate measurement due to the following two shortcomings. First, the pre-trained segmentor is vulnerable to artifacts that are far away from both pathological and healthy appearances. For example, Fig. \ref{fig62} shows two counterfeit images with different artifacts that are clearly abnormal. However, the pre-trained segmentor cannot recognize the abnormalities, resulting in the false high "Healthiness". Second, this metric \textcolor{black}{drastically} fluctuates at multiple trials. To verify this, we repeatedly pre-train three segmentors and then use them to calculate the "Healthiness" on the same synthetic images. As shown in Fig. \ref{fig31}, the results fluctuate drastically at various epochs and runtimes, especially when the "Healthiness" is smaller (e.g., SI(GVS, 30)\footnote{SI(GVS, $a$) is the abbreviation of the "Synthetic images (GVS, $a\%$)", which represents the synthetic images generated by the GVS that is trained on $a\%$ of training data.}, SI(ANT-GAN, 100), and SI(ANT-GAN, 70)). Another phenomenon is that the performance of the SI(GVS, 70) will surpass the SI(GVS, 100) at some epochs (e.g., $17^{th}$ epoch), which is unreasonable since the model trained on more data is superior to that trained on fewer data with a high probability. A similar phenomenon also appears in another pseudo-healthy synthesis method, ANT-GAN.

	{\color{black}\textit{To measure the healthiness accurately, we propose a new metric, A-Dice, by estimating the convergence speed of fitting an evaluation segmentor $S_E$  to synthetic images and corresponding lesion annotations.}} This metric is inspired by the study of label noise. Zhang et al. \cite{zhang2016understanding} revealed an intriguing phenomenon that the convergence time on false/noisy labels increases by a constant factor compared with that on true labels. Similar to this, aligning the well-transformed pixels (i.e., these pixels can be viewed as healthy ones) and the original lesion annotations is counter intuitive and hampers the fitting/convergence. To verify this, we assess the convergence speed by recording dice scores on the training data throughout the training process. The results on healthy images, the SI(GVS, 30) and SI(GVS, 100) are shown in Fig. \ref{fig32}. It is discovered that three evaluation segmentors finally attain similar dice values but have different convergence speeds. The evaluation segmentor $S_E$ trained on healthy images achieves the fastest convergence speed, followed by the SI(GVS, 30) and SI(GVS, 100).
	
	In the conference version, the $\mathbb{S}_{dice}$ was proposed to measure the healthiness by calculating the area under the dice curve. However, it may attain false results when synthetic images have similar healthy appearances since 1) the dice curves of synthetic images with different healthy appearances attain close convergence speed, which implies their lower discriminability, and 2) the performances at initial epochs are unstable due to the effects of random parameter initialization and batch selection, which results in the fluctuation of the $\mathbb{S}_{dice}$. To avoid false measurements, the first factor is considered, and the scheme that tries to \textit{enhance the discriminability} between dice curves is proposed. Concretely, we further utilize another property of false/noisy labels. That is, the higher learning rate will suppress the memorization ability of the DNN and prevent it from fitting labels \cite{Tanaka2018JointOF}. Improving the learning rate (i.e., from 0.0001 to 0.1) can obtain the dice curves presented in Fig. \ref{fig33}. We find that the gap of convergence speed between three types of images is significantly magnified. Thus, a new healthiness metric is proposed by evaluating the convergence speed of the evaluation segmentor $S_E$. The evaluation segmentor $S_E$ is trained on $lr=0.1$, and the dice values are recorded during the training process. Then, the A-Dice is calculated by averaging the dice values at multiple epochs, which is formulated as
	\begin{equation}
		\text{A-Dice} = \frac{1}{E}\sum_{e=1}^{E} dice_e,
	\end{equation}
	where $E$ denotes the total epochs and $dice_e$ represents the dice evaluated on the training data after the $e$-th epoch. \textit{Note that a lower A-Dice denotes faster convergence speed and further represents more healthy appearances.} Furthermore, compared with $\mathbb{S}_{dice}$, the A-Dice replaces the summation operation with the mean operation to eliminate the impact of the training epoch.
	
	We further emphasize that the proposed A-Dice is robust and stable. For the former one, any artifact distributionally different from the normal tissues can be quickly fitted to lesion annotations. As a result, it will be judged to be unhealthy due to the fast convergence. For the latter one, although the dice value at each epoch are unstable, the A-Dice can reduce the unreliability by averaging the dice values at multi epochs.
	
	We also design four experiments to validate the reliability of the proposed A-Dice from multiple views. {\color{black}Firstly, the two artifacts mentioned in Fig. \ref{fig6} are injected into the pathological images and then their A-Dice are evaluated. The green boxes in Fig. \ref{fig34} show that their A-Dice values are little affected and even become larger. Thus, the proposed A-Dice  \textit{effectively identifies the pathology that slightly deviates from the true pathology but is apparently abnormal.}} Secondly, the A-Dice values of ten types of images are presented in Fig. \ref{fig34}. The largest fluctuation range of the A-Dice is about 0.05 and is significantly less than the "Healthiness" \cite{xia2020pseudo} and the subjective assessment (rf. Table \ref{tab2}), which \textit{implies its good repeatability.} Thirdly, by comparing the relative relationship between different types of images, \textit{the A-Dice accurately reflects the relative order of them.} That is, the A-Dice of SI(Method, a) is smaller than that of SI(Method, b) when $a>b$ and Method $= [\text{GVS}, \text{ANT-GAN}]$. Fourthly, we find that \textit{the A-Dice agrees with the subjective assessment to a certain extent.} That is, the PHS-GAN and ANT-GAN achieve close results for the A-Dice (0.607 and 0.618, rf. Table \ref{tab1}), which also happens to the subjective healthiness metric (3.800 and 3.803, rf. Table \ref{tab2}).
	
	\section{Experiments} \label{sec5}
	\subsection{Datasets}
	The proposed GVS is validated on two widely used public datasets: Multimodal Brain Tumor Segmentation Challenge 2019 dataset (BraTS19) \cite{bakas2017advancing,menze2014multimodal,bakas2018identifying} and Liver Tumor Segmentation Challenge dataset (LiTS) \cite{bilic2019liver}.
	
	\noindent\textbf{BraTS.} The first validation dataset is the BraTS consisting of 259 GBM (i.e., glioblastoma)  and 76 LGG (i.e., lower-grade glioma) volumes that have been skull-stripped, interpolated to an isotropic spacing of $1mm^3$ and co-registered to the same anatomical template. Each volume includes 4 modalities (i.e., T1, T2, T1c, and Flair), and each slice is $240\times240$. The T1 and T2 modalities of GBM are utilized and split into training (234 volumes) and test sets (25 volumes). For each volume,  the intensities are clipped to $[0, V_{99.5}]$, where $V_{99.5}$ is the $99.5\%$ largest pixel value of the corresponding volume \cite{li2018h}.
	
	\noindent \textbf{LiTS.} We use the training data set of LiTS, which contains 131 CT scans of the liver acquired from 7 different clinical institutions. The resolution of the slice is $512\times512$.  The dataset is divided into training (118 scans) and test sets (13 scans). Following Dou et al. \cite{dou20173d}, the image intensity values of all scans are truncated to the range of $[-200,250]$ to remove the irrelevant details.
	
	\subsection{Implementation details and Baselines}
	\noindent\textbf{Implementation details.} The generator and the segmentor adopt the encoder-decoder \cite{Johnson2016PerceptualLF}  and U-Net \cite{ronneberger2015u} architectures, respectively. The proposed method is implemented in PyTorch. The models are trained with the Adam optimizer. The learning rate is set to 0.001, and the $E$ is set to $20$. The batch size is set to 8 for BraTS and 2 for LiTS. The $\lambda$ is set to $10.0$.  The training is implemented using an NVIDIA TITAN XP GPU.
	
	\noindent\textbf{Baselines.} The proposed method is compared with three approaches, VA-GAN \cite{baumgartner2018visual}, PHS-GAN \cite{xia2020pseudo}, and ANT-GAN \cite{sun2020adversarial}. \textit{The VA-GAN utilizes pathological images along with image-level annotations, whereas both the PHS-GAN and the ANT-GAN further utilize pathological images along with pixel-level annotations.} For the VA-GAN and PHS-GAN, we use the official code (i.e., VA-GAN\footnote{https://github.com/baumgach/vagan-code} and PHS-GAN\footnote{https://github.com/xiat0616/pseudo-healthy-synthesis}) and train their architectures on our dataset. Besides, the ANT-GAN is implemented based on the code provided by the authors.
	
	\subsection{Other evaluation metrics}
	To comprehensively assess the effectiveness of our method, the synthetic images are evaluated objectively and subjectively.
	
	\noindent\textbf{Objective metrics.} The overall quality of pseudo-healthy images can be measured from two aspects, healthiness and subject identity.  In Sec. \ref{sec4}, the proposed A-Dice has been described, which can properly assess the healthiness of pseudo-healthy images. \textcolor{black}{This section further introduces the process to measure the subject identity}, which can be expressed as calculating the visual similarity on the healthy regions. In practice, the MPSNR (masked PSNR) and MSSIM (masked SSIM) utilized to measure the subject identity are defined as.
	\begin{equation}
		\text{MPSNR} = \text{PSNR}[(1-y_t) \odot \mathbf{G}(x_p), (1-y_t) \odot x_p],
	\end{equation}
	\begin{equation}
		\text{MSSIM} = \text{SSIM}[(1-y_t) \odot \mathbf{G}(x_p), (1-y_t) \odot x_p],
	\end{equation}
	where $y_t$ is the lesion annotation while PSNR() and SSIM() denote Peak Signal-to-Noise Ratio and Multi-Scale Structural Similarity Index, respectively. Both PSNR and SSIM are adopted since they both are widely used to measure the similarity between two images and mainly differ on their degrees of sensitivity to image distortion \cite{hore2010image}.
	
	\noindent\textbf{Subjective metrics.}  The subjective metric is the gold standard to evaluate the quality of pseudo-healthy images. Here, we adopt the human evaluation method similar to the reference \cite{xia2020pseudo} \textcolor{black}{is adopted}, which is composed of the factors of healthiness and subject identity. The detailed process is presented as follows.
	
	We randomly select 100 synthetic outputs from each comparison method and arrange them as follows. Except for the pathological input placed in the first position, the synthetic outputs of all comparison methods are randomly arranged in the next four positions. Finally, the last position is occupied by the lesion annotation to better convey the pathological information to raters. Meanwhile, the raters are blinded to \textcolor{black}{the algorithm that generated each image}. The subjective ratings are rated on a five-level scale: 5, 4, 3, 2, and 1. Three medical image analysis researchers are asked to independently score each synthetic image from the aspects of healthiness (i.e., the scores from 5 to 1 represent that the healthiness declines sequentially) and subject identity (i.e., the scores from 5 to 1 represent that the subject identity is erased sequentially). {\color{black}Moreover, the healthiness further can be judged from the degree of achieving harmony between healthy and pathological regions in the synthetic images.  The criteria of subjective identity include the brightness, contrast, and the degree of keeping the detailed tissues.}

	\begin{figure}[!h]
		\centering
		\includegraphics[width=0.9\columnwidth]{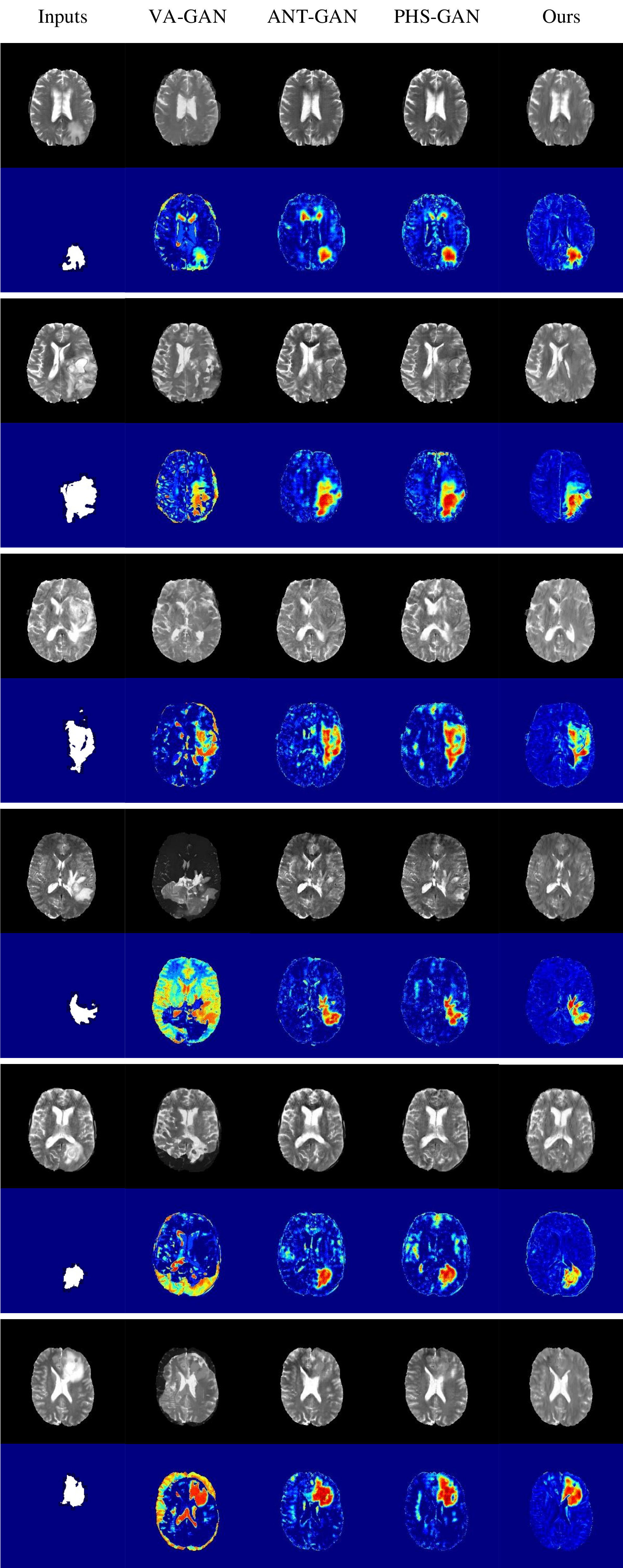}
		\caption{Example results on the BraTS T2 modality. We plot six examples (blocks) from top to bottom. In each block, the first column shows the input and the lesion annotation, and the next four columns show the synthetic images and the difference maps generated by the VA-GAN, ANT-GAN,  PHS-GAN and  GVS, respectively.}
		\label{fig7}
	\end{figure}
	
	\begin{figure*}[!h]
		\centering
		\subfigure[]{\includegraphics[width=0.25\textwidth]{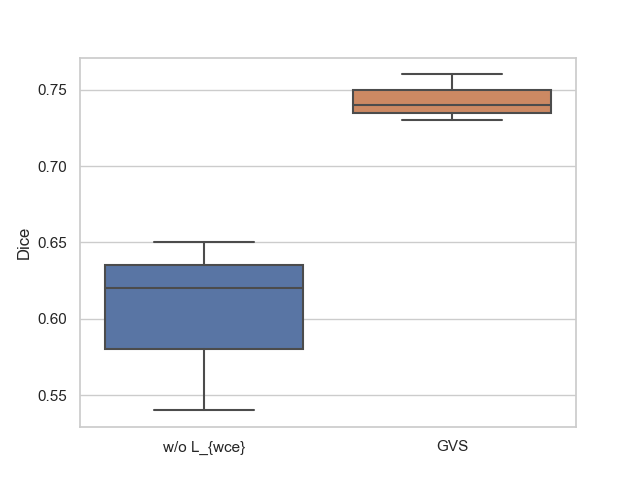}\label{fig91}}\subfigure[]{\includegraphics[width=0.25\textwidth]{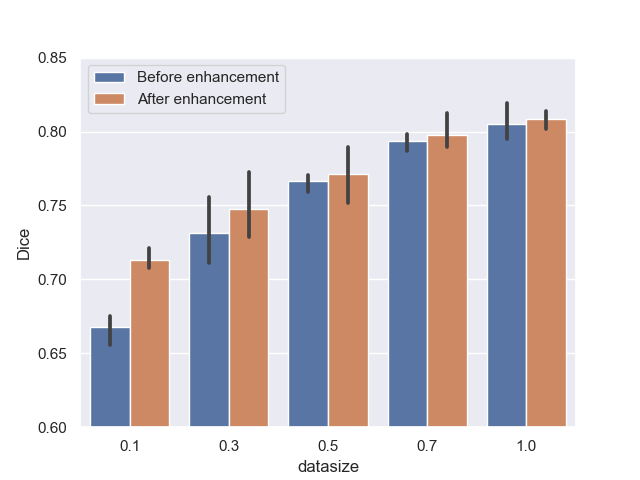}\label{fig92}}\subfigure[]{\includegraphics[width=0.25\textwidth]{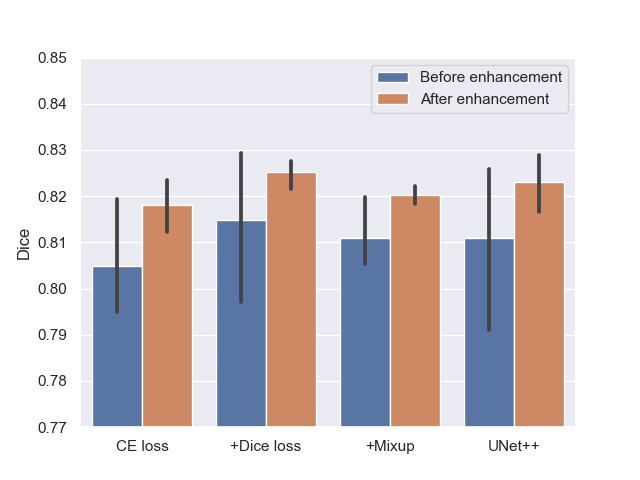}\label{fig93}}\subfigure[]{\includegraphics[width=0.25\textwidth]{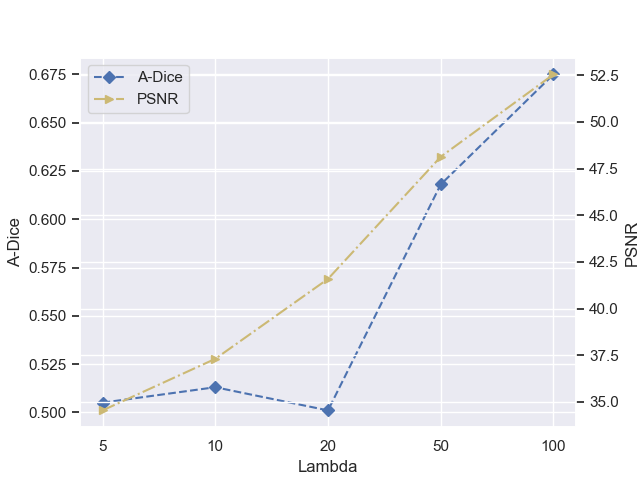}\label{fig94}}
		\caption{{\color{black}(a) The generalization ability of the segmentor in the GVS. Adding the $\mathcal{L}_{wce}$ effectively improves the generalization of the segmentor. (b) The variety of segmentation performance before and after enhancement. The proposed enhancement technique effectively improves the segmentation performance, especially in the low data regimes. (c) The variety of segmentation performance when combined with Dice loss, mixup, and UNet++. Combining these techniques and our enhancement method further improves the performance. (d) The sensitivity analysis for $\lambda$.}}
		\label{fig9}
	\end{figure*}

	\subsection{Comparison with the State-of-the-art Methods} \label{sec55}
	\noindent\textbf{Qualitative results.} The qualitative results are shown in Fig. \ref{fig7}. The effectiveness of all comparison methods is judged from the aspects of the subject identity and healthiness.
	
	The healthiness can be judged by comparing whether pathological and normal regions are harmonious. If the pathological regions are in harmony with the normal regions in the synthetic images, such images are healthy. On the contrary, \textcolor{black}{if} the pathological regions can be easily distinguished from the normal ones in the synthetic images, such images are viewed as being "not healthy". The performance of the VA-GAN fluctuates substantially. Some of the synthetic images have promising performance (e.g., the first and third examples in Fig. \ref{fig7}). However, the \textcolor{black}{majority} of them can be easily distinguished from the healthy images due to poor reconstruction. The PHS-GAN and the ANT-GAN remove most lesions, but some artifacts still remain (rf. the first, second, fourth, and sixth samples in Fig. \ref{fig7}). Finally, the proposed GVS removes more lesions than the others and replaces the lesion regions with a relatively healthy-looking area.
	
	The subject identity is determined by comparing the inputs and the synthetic images in terms of \textit{structural details, brightness, and contrast}. Since the ANT-GAN, PHS-GAN, and the proposed GVS reconstruct the normal tissue well and \textcolor{black}{have slight differences in reconstructions}, we plot the difference maps between the inputs and the synthetic images to magnify their differences. Combining difference maps and labels, the proposed method shows \textcolor{black}{higher quality} reconstructions than the other methods due to preserving more details of the brain tissues. It should be noted that the cerebrospinal fluid has pixels with high intensities and is close to lesions. These regions are weakened to varying degrees in the other methods, whereas they are well-preserved in the proposed method. Overall, the VA-GAN cannot keep the subject identity and loses a part of the lesion region in some cases. The PHS-GAN and ANT-GAN preserve the brain region but lose some details. The proposed method achieves the best subject identity among all methods.
	
	\begin{table}
		\centering
		\caption{Quantitative results on BraTS T2 modality. We report the average value and standard deviation of 3 trials. GVS($a\%$ data) represents the GVS trained on $a\%$ of training data.}	
		\begin{tabular}{llll}
			\toprule
			Method & MPSNR $\uparrow$ & MSSIM $\uparrow$ & A-Dice $\downarrow$ \\
			\midrule
			Original images & - & - & 0.736($\pm$0.310) \\
			VA-GAN \cite{baumgartner2018visual}  & 21.89($\pm$1.02) & 0.742($\pm$0.470) & - \\
			PHS-GAN \cite{xia2020pseudo} & 29.51($\pm$0.47) & 0.966($\pm$0.025) & 0.607($\pm$0.056) \\
			ANT-GAN \cite{sun2020adversarial} & 28.75($\pm$0.51) & 0.963($\pm$0.018) & 0.618($\pm$0.066) \\
			GVS($30\%$ data) & \textbf{38.79($\pm$0.39)} & \textbf{0.995($\pm$0.013)} & 0.615($\pm$0.040) \\
			GVS w/o $\mathcal{L}_{wce}$ &  37.31($\pm$0.34) & 0.991($\pm$0.012) & 0.549($\pm$0.045) \\
			\rowcolor{gray!20} GVS & 37.31($\pm$0.34) & 0.991($\pm$0.012) & \textbf{0.512($\pm$0.035)} \\
			\bottomrule
		\end{tabular}
		\label{tab1}
	\end{table}	
	
	\noindent\textbf{Objective evaluation.} The quantitative results on the BraTS dataset are shown in Table \ref{tab1}. The results of MSSIM and MPSNR show that all methods significantly improve the subject identity compared with the VA-GAN. The reason is that the other methods utilize the pixel-level lesion annotations, which is important for keeping normal regions and transforming pathological regions.  The proposed method further improves the visual similarity compared with the PHS-GAN and ANT-GAN. Otherwise, since the VA-GAN cannot reconstruct the normal tissues well (see Fig. \ref{fig7} and Table \ref{tab1}), its A-Dice value is meaningless and is not considered. The A-Dice values of the ANT-GAN and the PHS-GAN are 0.618 and 0.607. Compared to the original images, they decline by 0.118 and 0.129, respectively. Moreover, the A-Dice value of the proposed GVS is 0.512, with a decline of 0.224. The proposed method attains significant improvement compared with the existing methods. Even, trained on only 30\% of the training data, the GVS achieves notable performance. Specifically, the A-Dice is close to the ANT-GAN and PHS-GAN while both the MSSIM and MPSNR significantly exceed them.
	
	\noindent\textbf{Subjective assessment.} {\color{black}We report the subjective results in Table \ref{tab2}. The overall tendency of subjective results is similar to that of objective results. That is, the GVS shows the best performance in both "HEALTHINESS" and "IDENTITY" (i.e., subjective assessment for healthiness and identity. All capital for avoiding confusion.). Subsequently, the PHS-GAN and the ANT-GAN have similar results, while the VA-GAN occupies the bottom position. Particularly, the performance differences are magnified, and our GVS has better performance than the other methods in the aspects of both "HEALTHINESS" (GVS (4.457) vs ANT-GAN (3.707)) and "IDENTITY" (GVS (4.390) vs ANT-GAN (3.803)). Furthermore, our GVS achieves near-ideal performance ("HEALTHINESS": GVS (4.457) vs upper bound (5.0); "IDENTITY": GVS (4.390) vs upper bound (5.0)), which suggests the near-perfect performance of the GVS from the expert perspective.}
	
	\subsection{Effectiveness of difference-aware loss} \label{sec56}
	Sec. \ref{subsection:segmentor} states that the segmentor in the basic GVS has a poor generalization ability due to the mismatch between synthetic images and lesion annotations. Thus, the generator will be misguided and further harm the synthetic quality. To verify this, we first evaluate the generalization ability of the segmentor before and after adding difference-aware loss. Specifically, the pathological images are sent to two segmentors trained by the GVS and the GVS w/o difference-aware loss, respectively, and then the corresponding dice scores are calculated. The results shown in Fig. \ref{fig91} illuminate that the proposed GVS achieves a higher average dice score and lower variance compared to the GVS w/o $\mathcal{L}_{wce}$, which confirms that the difference-aware loss effectively improves the generalization ability of the segmentor.
	
	\begin{table}
		\centering
		\caption{Subjective results on the BraTS T2 modality. We report the average value and standard deviation of 3 raters.}	
		\begin{tabular}{lcc}
			\toprule
			Method & "HEALTHINESS" $\uparrow$ & "IDENTITY" $\uparrow$ \\
			\midrule
			VA-GAN \cite{baumgartner2018visual}  & 2.103($\pm$0.414) & 1.943($\pm$0.519) \\
			PHS-GAN \cite{xia2020pseudo} & 3.667($\pm$0.824) & 3.800($\pm$0.568) \\
			ANT-GAN \cite{sun2020adversarial} & 3.707($\pm$0.823) & 3.803($\pm$0.518) \\
			\rowcolor{gray!20} Our GVS & \textbf{4.457($\pm$0.330)} & \textbf{4.390($\pm$0.491)} \\
			\bottomrule
		\end{tabular}
		\label{tab2}
	\end{table}
	
	\begin{figure*}[!h]
		\centering
		\includegraphics[width=0.85\textwidth]{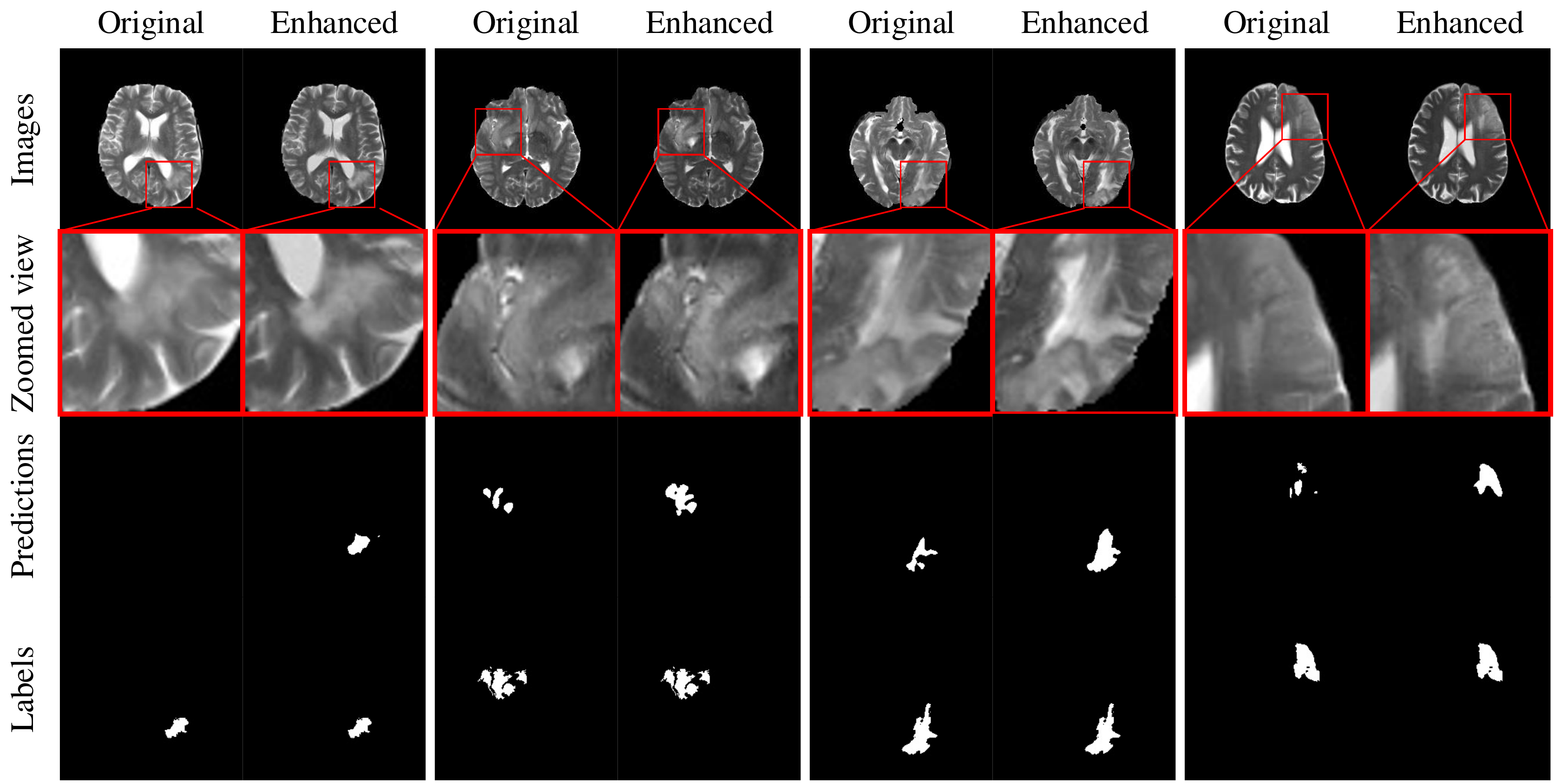}
		\caption{Examples of segmentation predictions for enhanced results on the BraTS dataset. These results are generated when $\alpha=1.0$.}
		\label{fig8}
	\end{figure*}

	\begin{table*}
		\centering
		\caption{Comparison of the segmentation performance when $\alpha = [0.0, 0.3, 0.5, 0.7, 1.0]$. All results are averaged over 3 trials. Note that $\alpha = 0.0$ represents the images without enhancement, and values in brackets are differences of segmentation performances before and after enhancement. Maximum improvement is denoted in \textbf{bold}, and the second one is denoted in \underline{underline}.}
		\begin{tabularx}{400pt}{c *7{>{\Centering}X}}\toprule
			\multirow{2}{*}{BraTS} & $\alpha$ & 0.0 & 0.3 & 0.5 & 0.7 & 1.0 \\
			\cmidrule(l){2-7}
			~ & Dice & 0.805 & \underline{0.813(+0.008)} & 0.812(+0.007) & \textbf{0.818(+0.013)} & 0.808(+0.003) \\
			\midrule
			\multirow{2}{*}{LiTS} & $\alpha$ & 0.0 & 0.3 & 0.5 & 0.7 & 1.0 \\
			\cmidrule(l){2-7}
			~ & Dice & 0.643 & 0.655(+0.012) & 0.653(+0.010) & \textbf{0.666(+0.023)} & \underline{0.665(+0.022)} \\
			\bottomrule
		\end{tabularx}
		\label{tab3}
	\end{table*}
	
	Next, we further test and verify the effectiveness of difference-aware loss on healthiness and subject identity. As shown in Table \ref{tab1}, the A-Dice has a significant improvement, while the MPSNR and MSSIM increase slightly, which suggests a tight relationship between the generalization ability of the segmentor and the healthiness of synthetic images. Accordingly, improving the generalization ability of the segmentor is a plausible way to improve the healthiness of synthetic images.
	

	\subsection{Results on contrast-enhanced brain tumor segmentation} \label{sec57}
	The segmentation performance after enhancement is shown in Table \ref{tab3}. It can be observed that under different \textcolor{black}{values of $\alpha$}, the segmentation performances improve to different extents. To further explore the reason why the proposed enhancement technique can quantitatively improve the segmentation performance, four examples with lower contrast are shown in Fig. \ref{fig8}. Intuitively, it is easier to distinguish tumors after enhancement. Furthermore, the tumors in the enhanced images are detected more accurately. These results manifest that our method effectively simplifies the segmentation by enhancing the contrast between normal and pathological regions.
	
	Next, we conduct the experiments to explore the improvements brought by the enhancement under different data sizes. The results are shown in Fig. \ref{fig92}, and we find that the improvements are more evident when the data size is smaller. We guess that this is because the network is easy to overfit when data size is small, and the enhancement explicitly regularizes the network at the image level so that better solutions are easier to be found.
	
	Lastly, we verify the compatibility of the proposed enhancement by combining it with other techniques (e.g., Dice loss,  mixup, and UNet++) that have proven effective for the segmentation. Specifically, the dice loss presented in V-Net \cite{Milletari2016VNetFC} is an effective method to address the class imbalance problem existing in medical image segmentation. Mixup \cite{zhang2017mixup} is an effective data augmentation technique in the classification task. Recently, it has been introduced in medical image segmentation \cite{eaton2018improving}. The U-Net++ \cite{zhou2018unet++} is a variant of U-Net based on nested and dense skip connections, and its effectiveness was verified on multiple medical image segmentation tasks. The results are shown in Fig. \ref{fig93}. The above-mentioned three techniques indeed improve the segmentation performance in brain tumor segmentation, and our enhancement technique further improves the segmentation performance on their basis. On the contrary, these three techniques also further improve the segmentation performance based on our enhancement technique.

	\subsection{Sensitivity analysis of $\lambda$} \label{sec58}
	The proposed GVS only contains one hyperparameter, $\lambda$, which is used to balance the power of the visual residual loss  $\mathcal{L}_R$ and adversarial loss $\mathcal{L}_{s2}$ when training the generator $G$. Here, the sensitivity of GVS for different choices of $\lambda$ is investigated and, the results are shown in Fig. \ref{fig94}. We discover that when $\lambda \in [5, 20]$, the healthiness varies slightly and is insensitive to the value of $\lambda$. Moreover, the healthiness and subject identity take up the opposite position. Better healthiness (lower A-Dice value) means worse subject identity and \textcolor{black}{vice versa}.

	\subsection{Extensive results on other modalities} \label{sec59}
\begin{figure*}[!h]
		\centering
		\includegraphics[width=0.85\linewidth]{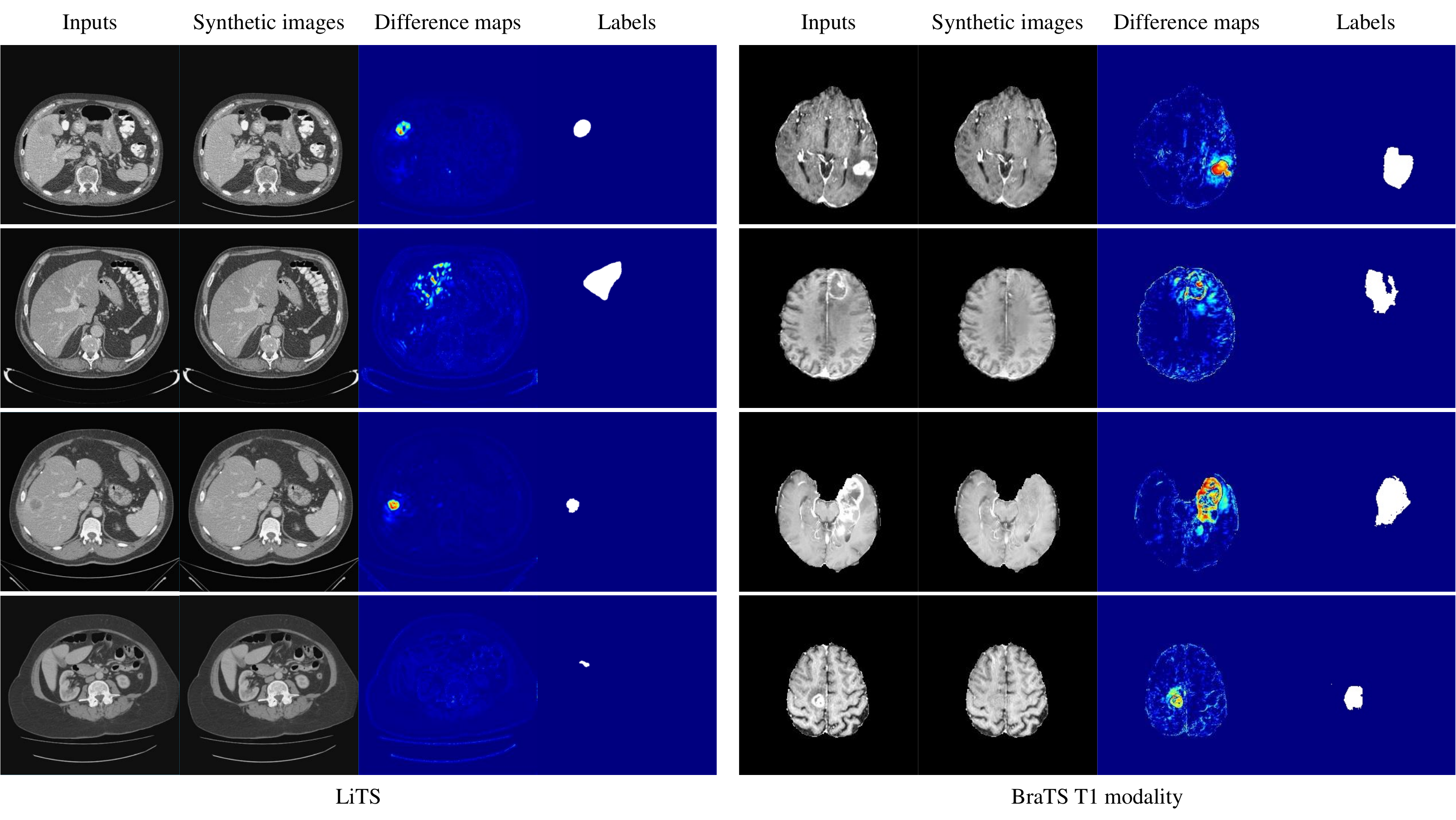}
		\caption{Example results on the LiTS (left) and BraTS T1 modality (right). For each modality, we plot four examples from top to bottom. For each quaternion, we show  the input, synthetic image, difference map, and label from left to right.}
		\label{fig10}
	\end{figure*}

	The proposed GVS is also evaluated on other modalities including CT and MR T1. The results of one CT dataset, LiTS, are shown on the left side of Fig. \ref{fig10}. The top row shows that the proposed method effectively maintains the subject identity and transforms the high-contrast lesions into healthy-looking tissues well. \textcolor{black}{However,} the results corresponding to the images with lower contrast are not satisfactory, as shown in the bottom row. This is because the images in the LiTS dataset have higher resolution, and even more importantly, the lesions have smaller volumes and lower contrast. The right side of Fig. \ref{fig10} shows the results on the BraTS T1 modality, which are similar to that on the T2 modality. \textcolor{black}{The results of MPSNR=38.34, MSSIM=0.994, and A-Dice=0.501 also confirm the excellent performance of the proposed GVS.}
	
	The synthetic results are also applied to image enhancement to assist lesion segmentation, and the results are shown in Table \ref{tab3}. \textcolor{black}{It is observed} that the performance improvement on the LiTS is more significant than that on the BraTS. We conjuncture that it's owing to the proposed enhancement technique facilitates the segmentation more significantly when it is applied to lower-contrast images. The phenomenon that the higher $\alpha$ results in more improvement further strengthens this view.
	
	\section{Discussion and Conclusion} \label{sec6}
	This work consist of three parts, involving an effective pseudo-healthy synthesis framework, a meaningful application, and a reliable evaluation metric.
	
	The first one is an adversarial framework consisting of a generator and a segmentor proposed to cope with the problem of pseudo-healthy synthesis. The experiments show that our GVS achieves excellent performance compared to the state-of-the-art methods. Besides the results and the analysis in Sec. \ref{sec55}, we further emphasize that the healthiness and the subject identity are competing (shown in Sec. \ref{sec58}). A similar attribute, the contradiction between reconstruction and adversarial losses, is also verified in the GAN \cite{ramasinghe2020conditional}. Hence, achieving superior performance in both healthiness and subject identity for our GVS is promising.
	
	{\color{black} Fig. \ref{fig7} and \ref{fig10} show that our GVS fills pathological regions with diverse healthy-looking tissues, which implies the reduction of the mode collapse. The research \cite{tran2019improving,arjovsky2017wasserstein} demonstrated that the higher dimension will exacerbate the mode collapse and meanwhile need more training samples. In the GANs, one image is a sample, and its distribution is high-dimensional so that generating such distribution is easy to fall into collapse. Our GVS treats one pixel as a sample. The dimension of its distribution is significantly reduced. Thus, the mode collapse will be effectively alleviated.}
	
	Although our method achieves impressive results, it is confined to densely labeled annotations. In clinical applications, it is extremely difficult to collect huge amounts of accurate segmentation labels. Hence, it is necessary to relax the demand for accurate pixel-level annotations (e.g., semi-supervised learning, weakly-supervised learning) in the next step.
	
	Sec. \ref{sec56} reveals the close relationship between the power of the segmentor and the healthiness of synthetic images. It is found that strengthening the segmentor contributes to further improving the capacity of the GVS. One way to achieve this goal is to adopt more powerful architectures.  Hence, we plan to upgrade the GVS to 3D structures to further improve the synthetic performance. Furthermore, it would also be worth exploring if  the GVS could improve the segmentation performance of the  segmentor in three dimensions.
	
	Furthermore, we design the difference-aware loss to alleviate the poor generalization ability of the segmentor, which is verified to be effective for improving the healthiness of synthetic images (rf. Sec. \ref{sec56}). However, this design still has room for improvement. For example, we do not consider that at first epochs the synthesis is imperfect and the difference maps are less meaningful.   Hence, \textcolor{black}{improving the difference-aware loss or designing} a novel scheme to enhance the generalization of the segmentor is our next target.
	
	The second part of the main work is using difference maps to enhance the contrast between normal tissues and lesions. The results in Sec. \ref{sec57} and \ref{sec59} illustrate that the proposed enhancement technique effectively alleviates the low contrast problem and improves the segmentation performance. Moreover, this technique is orthogonal to other techniques that also contribute to segmentation. More importantly, the segmentor trained on less training data benefits more from the enhancement is an interesting phenomenon. It is well known that training a model with small datasets is a critical and challenging topic in medical image segmentation \cite{zhang2019survey}. The GVS provides an effective solution to deal with this problem and merits further exploration.
	
	The third part is proposing a reliable metric, A-Dice, to measure the healthiness of synthetic images. The proposed metric also has the potential for wider applications. For example, we plan to use this metric to measure the abnormal information in the pathological images and explore the relationship between it and downstream tasks, such as tumor grading and survival prediction.

	\bibliographystyle{IEEEtran}
	\bibliography{GVS}
	
\end{document}